# Dynamic complex opto-magnetic holography

M. Makowski[1]*, J. Bomba[2], A. Frej[2], M. Kolodziejczyk[2], M. Sypek[1], T. Shimobaba[3], T. Ito[3], A. Kirilyuk[4], and A. Stupakiewicz[2]*

[1]Faculty of Physics, Warsaw University of Technology, 75 Koszykowa, 00-662 Warsaw, Poland.
[2]Faculty of Physics, University of Bialystok, 1L Ciolkowskiego, 15-245 Bialystok, Poland.
[3]Graduate School of Engineering, Chiba University, 1-33 Yayoi, Inage, Chiba 263-8522, Japan.
[4]FELIX Laboratory, Radboud University, 7 Toernooiveld, 6525 ED Nijmegen, The Netherlands.

*Correspondence to: michal.makowski@pw.edu.pl (M.M.); and@uwb.edu.pl (A.S.).

**Computer-generated holograms with their animated, three-dimensional appearance have long appealed to our imagination as the path towards truly immersive displays with bi-directional natural parallax. Impressive progress in updateable 3-D imagery[1] has been achieved with liquid crystal modulators[2] and high-resolution, but quasi-static holograms are being recorded in photosensitive materials[3-5]. However, the memory requirements and computational loads of real-time, large-area holography will be hard to tackle for several decades[6] to come with the current paradigm based on a matrix calculations and bit-plane writing. Here, we experimentally demonstrate a conceptually novel, holistic approach to serial computation and repeatable writing of computer-generated dynamic holograms without Fourier transform, using minimal amounts of computer memory. We use the ultrafast opto-magnetic recording of holographic patterns in a ferrimagnetic film with femtosecond laser pulses, driven by on-the-fly hardware computation of a single holographic point. The intensity-threshold nature of the magnetic medium allows sub-diffraction-limited, point-by-point toggling of arbitrarily localized magnetic spots on the sample, according to the proposed circular detour-phase encoding, providing complex modulation and symmetrical suppression of upper diffractive orders and conjugated terms in holographically reconstructed 3-D images.**

Ultimate large-scale holographic displays based on computer-generated holography (CGH) will require an exotic light modulation device, capable of rewriting as many as $10^{12}$ of light-diffracting cells[6] at a minimal framerate of 180 Hz, assuming color field sequential operation. To date, no large-area material or method having such properties has been demonstrated. Liquid crystal on silicon[7] spatial light modulators[8] (LCoS SLM) offer short refresh times in the order of microseconds in the case of ferroelectric LC[9]. However, their spatio-temporal product[6] is restricted by the fundamental limitations of LC and ineffective parallel data transfer scheme of pixel driving circuitry with time-consuming row-column serial addressing. Using multiple SLMs in both coherent[10] and incoherent matrices[11] partly overcomes this trade-off by offering e.g. 133 megapixels from 16 SLMs[11], but with a penalty of increased complexity[12,13] and image inconsistency. Photorefractive and photochromic materials[4,14] allow large-surface, ultra-dense, pixel-less writing. However, their long response times in milliseconds and relatively long nanosecond pump pulses with fluences[3] of 650 mJ·cm$^{-2}$ effectively render them quasistatic. From the standpoint of scalability, the ability to achieve dense writing on large areas without any in-situ electrodes is essential. This is feasible with two interfering beams[15] at μs speeds. However, in addition to the bulky optical setup and high voltages involved[16], this does not ensure angular flexibility in the positioning of writing areas, unless mechanical sample movements are involved[3]. Moreover, owing to the sine profile of the recorded fringes, stray orders of diffraction are formed in the replay images. In contrast, single-beam point-by-point writing has greater flexibility and scalability. Nevertheless, micrometer positioning and localization of spots on a sample typically requires extremely large numerical apertures[17].

Apart from the optical and material challenges, computing a CGH as a whole in the form of a large matrix of complex numbers is inevitably constrained by the available random-access memory (RAM) and by limited capabilities to compute two-dimensional Fast Fourier Transforms (FFT), despite recent progress[18]. Instead, we propose to compute a single CGH point at a nanosecond time scale with a dedicated hardware FPGA (Field Programmable Gate Array) unit fed with data describing the spatial coordinates and intensities of the cloud point representation of the input 3-D scene[19]. The serial single-point computation is done in two stages (see Methods). First, the phase contributions of all point sources in the input scene are computed in parallel by separate circuits in the FPGA module in relation to the current CGH cell to be written, which is targeted by current angles of the micro-electro-mechanical (MEMS) mirror scanning the writing laser beam over the hologram (see Fig. 1a). Subsequently, the phase contributions are counted by the *popcount* unit and the final binary signal is delivered, triggering a single 35 fs pulse of the writing laser. Thus, instead of being saved in RAM, the computed CGH point is immediately written in the opto-magnetic GdFeCo film (see Methods and Extended Data Fig. 1) in the form of a locally reversible, all-optically switched (AOS) magnetic spot, as shown in Fig. 1. The mechanism of ultrafast AOS, independent of the laser polarization[20] in a large spectral range, allows a stable magnetic spot to be written and rewritten after an interval of only 30 ps [21]. This process, which is equivalent to storing data in a non-volatile memory at a picosecond time scale, enables ultrafast writing of subsequent CGH points in other arbitrary positions on the sample. This feature differentiates this unique material from other non-volatile media, such as photopolymers that typically require exposure times greater by nine orders of magnitude[22].

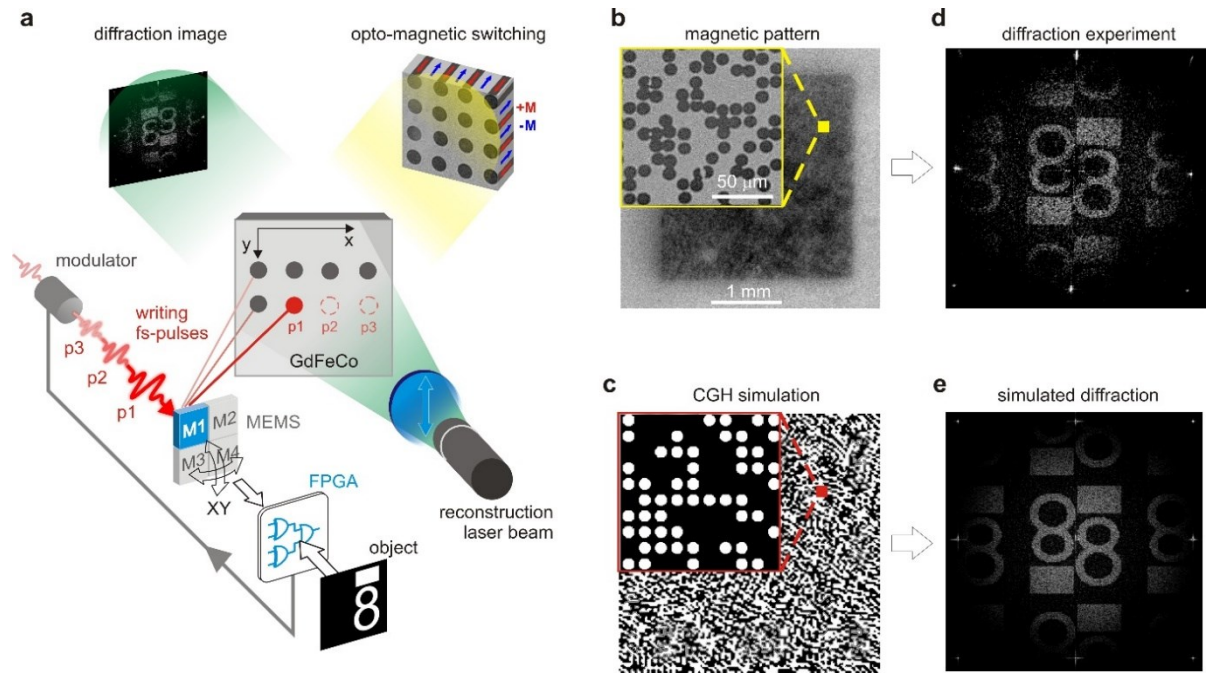

**Figure 1 | Serial writing of a CGH in an opto-magnetic medium. a,** Point-by-point optical switching of magnetic spots in GdFeCo sample with subsequent pulses *p*1-*p*3 from a femtosecond laser with a fluence of <20 mJ·cm$^{-2}$, reflected by a scanning MEMS mirror (M1), running free along raster-like or resonant (Lissajou) paths. M1 reports its current angles to the FPGA unit, which computes on-the-fly the binary triggering signal for the intensity modulator of the femtosecond laser beam; the FPGA unit is fed with cloud-point representation of the input object; geometrical imaging of the sample with a white polarized light using magneto-optical Faraday geometry; far field diffraction image observed on a camera after passing of the 532 nm reconstruction laser beam through the sample, an analyzer and a focusing lens. **b, c,** Magnetic pattern written and simulated in the sample comprising 256×256 CGH points; **d, e,** experimental and simulated far-field diffraction images showing the reconstructed image of the input object accompanied by twin conjugate image due to binary amplitude modulation. Partial suppression of upper order images is observed due to minor randomization of positions of CGH points in the sample[23].

Magnetic spot switching in the GdFeCo sample occurs above a certain intensity threshold with a size dependent on the intensity of the femtosecond laser pulse. Because a symmetric Gaussian beam is used for writing, well defined circular areas are switched (see Fig. 1a and Extended Data Fig. 2) with diameters falling below the diffraction limit[21,24]. This feature allows the dense packing of highly localized CGH spots. Reversible point-by-point writing of holographic patterns by phase change of $Ge_2Sb_2Te_5$ was reported[25] with a numerical aperture (*NA*) of 0.8. In contrast, in the proposed opto-magnetic method uses low *NA* of 0.01 that is supported by the said threshold effect, allowing convenient recording of CGH spots with sizes below $d$ = 10 μm at a working distance of 100 mm (see Methods). The speed of femtosecond laser beam scanning can considerably exceed the inertial limits of MEMS[26]; therefore, the pace of holographic recording is limited only by the laser pulse repetition rate and FPGA clock (availability up to 1 GHz). For instance, a gigahertz femtosecond laser potentially allows non-volatile rewriting of CGHs comprising 5.6 million points at 180 Hz framerate with the proposed method. This is superior to typical Full-HD LCoS SLMs that feature a similar framerate but having only 1920×1080 pixels (2.1 million).

Dynamic updating of optically written holographic frames can operate in three modes: (i) globally using a coil-induced external magnetic field; (ii) by all-optical serial restoring of all magnetic spots to the initial state (see Fig. 2a); and (iii) selectively by differential exposures (see Fig. 2b). The third mode comprises all-optical toggling[27] of the magnetic states limited to those CGH points selected by the logical XOR operation between *k* and *k*+1 states of the input 3-D scene. Although this approach doubles the number of required FPGA units, it results in an approximately two-fold reduction in the number of femtosecond laser exposures, potentially allowing the refreshing of a CGH comprising 2·5.6=11.2 million points at 180 Hz framerate. As a manifestation of the unique holographic feature, the complete images are reconstructed in far field diffractive plane while the point-by-point recording is still in progress, i.e. without strict division to subsequent holographic frames in time domain (see Methods and Extended Data Fig. 3). The opto-magnetic CGH recording throughput can be scaled up even further, without considerable limitations. This can be achieved by writing several areas of the sample simultaneously after adding more MEMS mirrors[13] (marked as M1-M4 in Fig. 1a), each with a dedicated FPGA computation unit and a writing laser beam. As an example, using four such units could allow the rewriting of CGHs comprising approximately 45 million points at 180 Hz, enabling color-sequential operation with unprecedented resolutions. In such a case, operating at low *NA* values greatly facilitates the focusing of multiple writing beams on the hologram. The picosecond switching cycles of holographic cells have the potential to be used as ultra-rapid reconfigurable transmissive diffractive optical elements. They could replace ferroelectric LCoS SLMs or Digital Micromirror Devices in selected cases of optical trapping[28], manipulation of optical vortices[29], orbital angular momentum[30,31], optical interconnections, multiplexers and switches[32], reconfigurable add-drop multiplexers[33] and electronic circuits[34]. For such purposes, useful sophisticated optical functionalities are achievable with CGHs containing as few as 128×128 cells that can be computed and updated 4 MHz rates with four parallel MEMS-FPGA units.

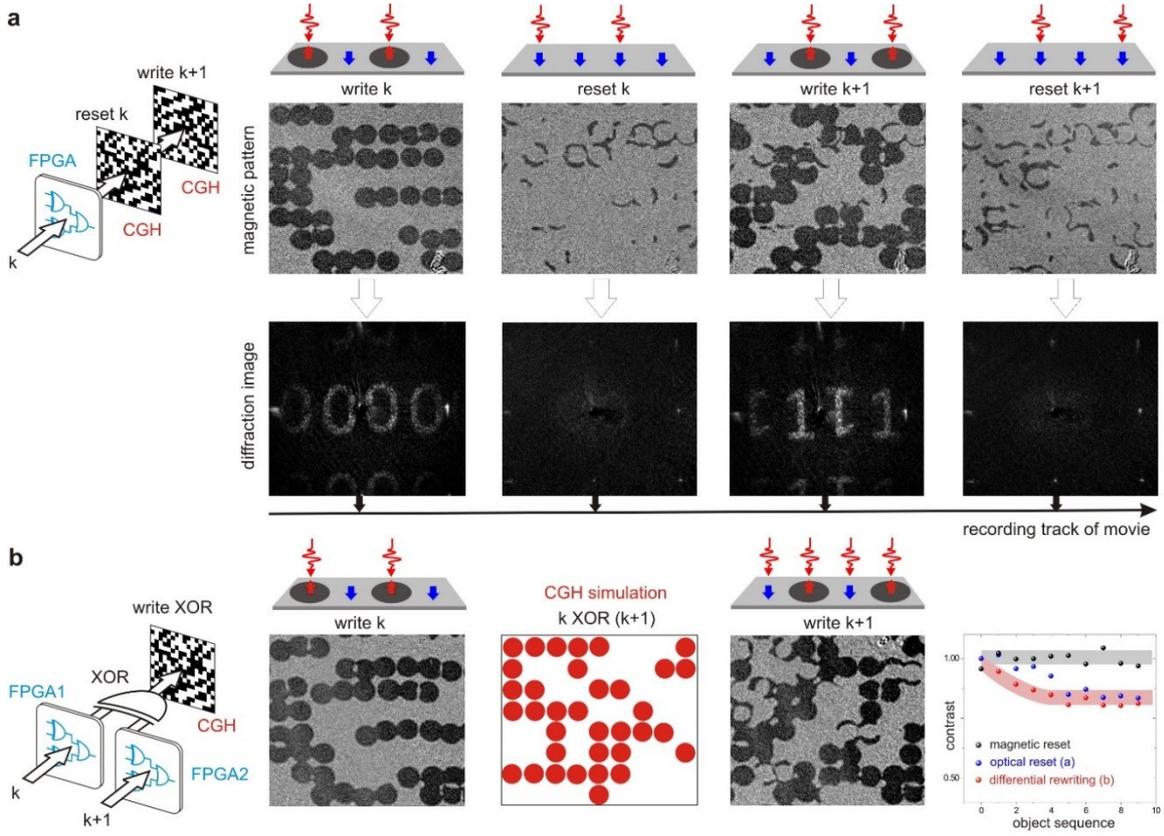

**Figure 2 | Dynamic opto-magnetic rewriting of subsequent holographic frames. a,** Optical reset mode comprising subsequent writing of *k* and *k*+1 holographic patterns using toggle regime of AOS. From left to right: writing of pattern *k* required 40 exposures (fragment of the real magnetic pattern is shown); reset stage toggled the magnetic spots to their initial state with extra 40 exposures; writing of next pattern *k*+1 required 34 exposures; second reset stage as a part of track of movie required another 34 exposures (148 exposures in total for re-writing). **b,** Differential rewriting mode. From left to right: computation scheme using two separate FPGA units; writing of *k* pattern required 40 exposures; computation of the logical XOR between *k* and *k*+1 frames; writing of *k*+1 frame required 35 exposures (the differential mode allowed an almost twofold, i.e. 148/(35+40) reduction of the number of required exposures); experimental contrast of reconstructed images (see Methods) as a function of rewriting sequences for all rewriting modes. The middle panel of diffraction images shows a movie of holographic frames (see Extended Data Movie S1). The visible remanent magnetic domains in images of the magnetic pattern after resetting are caused by long-lived motions of domain walls within milliseconds after AOS. This magnetostatic effect causing the reduction of the contrast of holographic playback frames becomes negligible after a sequence of four write-reset acts, as shown in the graph. The size of the shown magnetic patterns is 106×96 µm$^2$.

The drawback of the opto-magnetic medium is the binary modulation of the playback light that inevitably induces conjugate images in the diffraction field[35], as shown in Fig. 1. By introducing the Lohmann encoding[36] we effectively switched to complex (amplitude and phase) modulation while still operating in binary-amplitude medium. The non-pixelated nature and intensity-threshold AOS in the GdFeCo sample allows one to write in precise locations circularly shaped magnetic spots to form "openings" of Lohmann cells (see Fig. 3a). In the first attempt, we used the multiple AOS with constant diameters of magnetic spots to build approximations of the classical rectangular openings (see Fig. 3b). The experimental results demonstrate the elimination of conjugate terms and consistency with simulation, nevertheless the asymmetrical suppression of higher diffraction orders was also observed. In order to overcome this, we proposed the modified detour-phase encoding using single-shot circular openings (see Methods and Extended Data Fig. 4), which conveniently resulted in the symmetrical suppression[37] of higher order image duplicates in the far field, as shown in Fig. 3c. The diameter of each opening *d*(*I*) was adjusted by modulating the intensity of the pump beam on-the-fly (see Extended

Data Fig. 2), which differentiates this approach from previous serial writing attempts with fixed spot sizes[25]. In addition, this feature reduces the total number of exposures per holographic frame up to three times, as compared to asymmetric detour-phase encoding seen in Fig. 3b.

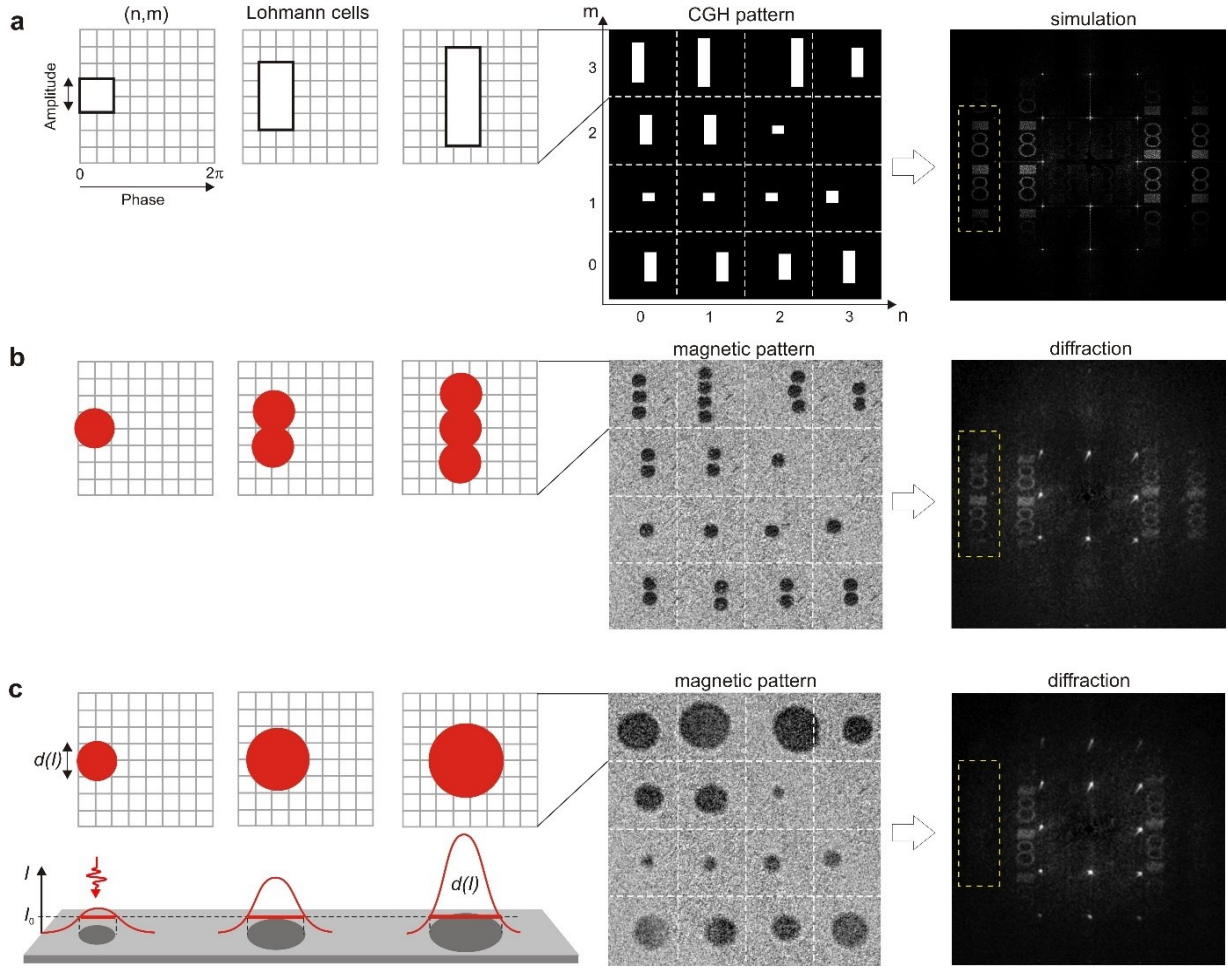

**Figure 3 | Complex opto-magnetic holograms with detour-phase encoding. a,** Simulation of classical detour-phase encoding using the Lohmann method. The heights and offsets of rectangular apertures are linear functions of the amplitude and phase values from a given CGH cell, respectively. Suppression of the conjugated images is notable. **b,** From left to right: the approximation of the rectangular Lohmann openings with multiple point-by-point exposures switching magnetic spots of fixed diameters; fragment of written magnetic pattern; experimentally acquired diffraction image showing asymmetrical visibility of higher-order image duplicates (marked by a dashed yellow rectangle) originating from regularly spaced Lohmann cells. **c,** From left to right: visualization of writing of three adjacent Lohmann cells with circular switched magnetic spots of diameters set by in-the-loop modulation of the pump beam intensity (see Fig. 1); fragment of written magnetic pattern; experimental reconstruction showing the symmetrical suppression of higher-order image duplicates. The size of magnetic pattern images is 320×320 µm$^2$.

The presented method operates at room temperature in non-contact mode using ultrashort laser pulses, which is greatly supported by the availability of all-fiber femto- and picosecond lasers operating within 1-10 GHz. It outperforms existing alternative approaches, enabling the fastest ever serial writing of CGH in non-pixelated, large-area, easily manufacturable, transparent (or reflective) medium as an updateable holographic non-volatile memory. Recent progress in material engineering that enables all-optical switching with longer picosecond pulses of low fluence, such as in dielectrics[38] or metallic [Tb/Co] multilayers[39] introduces a plethora of new possibilities. Furthermore, employing the proposed method in nanostructured media could be useful for volume multi-color holographic recording approaching sub-wavelength resolution[40]. These advantages combined with the scalable computation scheme and ultra-low requirements for computer memory open the way towards high-resolution holographic 3-D TV displays and ultra-wide angle near-eye augmented/virtual reality goggles[41].

**Acknowledgements**. We acknowledge support from the grant of the Foundation for Polish Science POIR.04.04.00-00-413C/17-00, from the Warsaw University of Technology within CB POB FOTECH-2 (Excellence Initiative: Research University - IDUB) and from the European Union's Horizon 2020 Research and Innovation Programme under the Marie Skłodowska-Curie grant agreement No 861300 (COMRAD). We thank A. Tsukamoto for providing the samples and C.S. Davies for fruitful discussions.

**Author Contributions.** M.M. and A.S. proposed the idea and conceived the project. J.B., M.K., A.F. performed the magneto-optical experiments and CGH writing. J.B., M.S. performed the simulations. T.S., T.I. designed FPGA circuits and performed the CGH computations. A.K. contributed to the data analysis. M.M. and A.S. jointly discussed the results and wrote the manuscript with contributions from all authors.

**Competing interests:** The authors declare no competing interests.

**Additional Information.** Correspondence and requests for materials should be addressed to M.M. (michal.makowski@pw.edu.pl) and A.S. (and@uwb.edu.pl).

## Methods

**Materials**

The measurements were performed on ferrimagnetic alloy with the composition AlTi(10 nm)/Si$_3$N$_4$(5 nm)/Gd$_{24}$Fe$_{66.5}$Co$_{9.5}$(20 nm)/Si$_3$N$_4$(60 nm), prepared by magnetron sputtering on a glass substrate. The sample was 15×15 mm$^2$ in size and optically transparent in the visible spectral range. The used alloy has perpendicular orientation of the magnetization and shows a square shape of the hysteresis loop with a coercive field of 75 Oe. Such alloy was used before for all-optical magnetic switching (AOS) with a single femtosecond laser pulse by toggle regime[20]. The AOS effect in GdFeCo alloys is driven by ultrafast and efficient thermal demagnetization at the characteristic time of about 30 ps after a single pump pulse. The mechanism of AOS in these samples has been previously intensively studied[21], it provides a unique medium for CGH recording of magnetic pattern with a spatial redistribution. We note that the magnetization switching does not depend on the pump polarization within the whole VIS-NIR spectral range.

**Recording of CGH magnetic patterns with femtosecond laser pulses**

We write and erase the magnetic patterns in the optically transparent GdFeCo film on a point-by-point basis with the use of MEMS-steered single linearly-polarized femtosecond pump pulses with a duration of 35 fs, the central wavelength of λ = 800 nm and a maximum repetition rate of 1 kHz (Astrella, Coherent). In principle, the repetition rate of the laser pulses can be scaled up to 10 GHz frequency, which is limited only by the mechanism of AOS in GdFeCo alloy. The pump beam with the fluence below 20 mJ·cm$^{-2}$ was focused to a spot of 50 μm in diameter using a lens of $f$=100 mm placed before the sample. The diameter of the pump beam at the lens was 2 mm, which is equivalent to the numerical aperture (*NA*) of 0.01. By lowering the pump beam intensity and taking advantage of the threshold effect, the optically switched areas of c.a. $d_s$=10 μm were achieved in the same optical configuration, which was equivalent to NA = 1.22·λ/(2·$d_s$) = 0.05.

The magnetic pattern in the sample was visualized using a standard magneto-optical polarizing microscope. The polarized light source was a LED lamp with output routed into the optical path of the probe beam. The LED light then passed the sample and was gathered with an objective before passing an analyzer and hitting the CCD camera. The magnetic contrast in such a polarizing microscope comes from the fact that magnetic spots with different perpendicular magnetization orientations to the sample plane will give different rotation of the polarization plane (effect of Faraday rotation), and thus the light passing through them will acquire different polarization, which can be easily detected on the CCD camera. After recording, the pattern stays unchanged for a long time due to the non-zero coercivity in the medium. All measurements were done without applied external magnetic fields and at room temperature.

The images of magnetic pattern and diffraction were taken before and after the CGH recording. We used the difference (standard procedure of magnetic image processing) of these images to visualize pure magnetic pattern in the magneto-optical images and subtract the zero-order light in the diffraction images (see Extended Data Fig. 1). The initial background image before CGH recording was obtained after the application of a brief external perpendicular magnetic field with >80 Oe. Such magnetic field was also used to erase the entire recorded CGH magnetic pattern. The diameter of the magnetic spots was adjusted by changing the pump intensity with a fixed lens position (see Extended Data Fig. 2). The pulse intensity was controlled by an electro-optical modulator.

The sequence of CGH recording is simple. The current angles of the free-running MEMS mirror are reported to the FPGA unit. Based on the output binary signal, a single laser pulse is transmitted through the intensity modulator and switches the magnetization in the medium. The procedure is repeated immediately after the next angles of the MEMS are available, allowing uninterrupted point-by-point recording (see Fig. 1). Another laser pulse directed at the same written magnetic spot reverses it, allowing for a rapid change of the pattern into another one, with the possibility of reusing unmodified points. We note that during increasing the CHG writing density due to the Gaussian shape of the laser pulse the overlapping effect of magnetic spots was observed[42]. In this case, there was observed the toggle effect in areas between subsequent written spots. Despite this, the quality of far field reconstructions of the written holographic patterns was unchanged. The overlapping effect could be useful for holographic writing of practically unlimited number of CGH points with arbitrary fill factor values. We also note that the serial mechanism of CGH writing allows the appearance of the recognizable holographic image during unfinished rewriting of the holographic frame. In Extended Data Fig. 3, we demonstrated the time trace of the displays diffraction image which is recognized for the minimal number of 32×32 points. This feature is observed only during serial writing and can be attributed to the non-volatile memory effect in the medium.

**Circular detour phase encoding**

The test holograms for complex encoding inspired by Lohmann method[36] were computed as Fourier transforms of the input images having 512×512 pixels, with a random initial phase. For each of the 512×512 assumed Lohmann cells the amplitude and phase values were extracted, denoted as $a(n,m)$ and $\varphi(n,m)$, respectively. The writing beam was then positioned inside a given cell, offset from its center by the distance:

$$\delta x(n,m) = \varphi(n,m) \frac{\Delta x}{2\pi}, \qquad (1)$$

where $\Delta x = \Delta y$ were the dimensions of each Lohmann cell. The diameter of the optically switched area $d(n,m)$ was directly connected with the amplitude $a(n,m)$ with the simple relation:

$$d(n,m) = \frac{b|a(n,m)|}{\Delta x}, \qquad (2)$$

where the $b=0.7$ margin was found in numerical optimizations in order to allow minimally invasive overlapping of Lohmann openings in adjacent cells[42]. The femtosecond writing beam had the gaussian intensity profile with waist diameter being a function of the used numerical aperture (NA) of the focusing optics. The appropriate intensity $I$ of the writing beam was adjusted to induce the magnetically switched circular area of the desired diameter $d$ (see Extended Data Fig. 2), by matching $d(I)=d(n,m)$ in a look-up table filled according to the below formula:

$$d(I) = 2\sqrt{-\frac{1}{2} w_s \ln\left(\frac{I_s - I}{I_s - I_0}\right)}, \qquad (3)$$

where the minimal (threshold) intensity of the beam inducing the smallest optical switching is equal to $I_0$, while the maximal intensity, causing the entire sample to optically switch (saturate) is equal to $I_S$. The NA-dependent waist diameter of the saturating writing beam is equal to $2w_S$ (measured at $1/e^2$ of the peak intensity $I_S$).

The rectangular openings in the Lohmann cells were replaced with circular ones, with the effect of symmetrical intensity envelope in far field holographic playback field, attenuating higher order terms (see Extended Data Fig. 4). Since the CGH plane ($x_1, y_1$) and far field playback plane ($x_2, y_2$) are in the Fourier relation, the attenuated reconstructed intensity field can be derived as:

$$\mathcal{F}\left\{\left[h(x_1,y_1) \cdot comb(x_1,y_1)\right] \otimes circ(x_1,y_1)\right\} = \left[H(x_2,y_2) \otimes comb(x_2,y_2)\right] \cdot J_1(x_2,y_2), \qquad (4)$$

where $h$ and $H$ are the CGH data and the reconstructed image, respectively, $comb$ function is the periodicity of the Lohmann cells, and $circ$ function denotes the circular shape of the openings used for CGH encoding. The resultant Bessel function $J_1$ is responsible for the symmetrical envelope, attenuating off-axis components in the playback signal.

**Point-by-point CGH computations for MEMS-driven opto-magnetic recording**

Each pixel of a binary CGH must be written to a magneto-optical device in the order of nanoseconds in order not to stop the mechanical motion of the addressing MEMS mirror. Since conventional processors have difficulty meeting this requirement, we have developed FPGA-based (Field Programmable Gate Array) CGH processors[18] that calculate a CGH point by the integration of contributions from all object points:

$$I(x_h, y_h) = \sum_{j=1}^{N} a_j \cos\left(\frac{\pi((x_h - x_j)^2 + (y_h - y_j)^2)}{z_j}\right), \qquad (5)$$

where $a_j$ denotes the amplitude of an object point, $(x_h, y_h)$ represents the coordinates of the computed CGH point, $(x_j, y_j, z_j)$ represents object points, $N$ denotes the total number of object points, and $I$ is the output signal denoting the desired intensity of the pump beam exposing the CGH point (which in binary case is equivalent to the desired transmittance of the CGH point). Those FPGA processors were designed to compute each CGH pixel in parallel, but the accumulation was done sequentially; thus, the calculation time is proportional to the number of object points. Therefore, it is not suitable for our purpose. For this opto-magnetic system, we have designed a new architecture of binary CGH processor. Equation (5) includes sequentially accumulation and the trigonometric function which consume hardware resources. To overcome this, we have derived a new CGH equation that can avoid these operations. To omit trigonometric functions, we first add the phase $\pi$ to each object point, which does not affect reconstructed images. We can rewrite Eq. (5) as:

$$I(x_h, y_h) = \sum_{j=1}^{N} a_j \sin\left(\frac{\pi((x_h - x_j)^2 + (y_h - y_j)^2)}{z_j}\right). \quad (6)$$

In binary CGH calculation, the final result has only binary values, so we can approximate this equation as:

$$I(x_h, y_h) = \sum_{j=1}^{N} a_j T\left(\frac{\pi((x_h - x_j)^2 + ((y_h - y_j)^2)}{z_j}\right), \quad (7)$$

where the thresholding function $T(x)$ outputs ones if $T(x) < \pi$, otherwise zeros, which can be readily implemented into FPGA. In order to further omit the sequential accumulation, we simply expand the accumulation as:

$$I(x_h, y_h) = b_1 + b_2 + \cdots + b_N. \quad (8)$$

where $b_n = a_n T\left(\frac{\pi((x_h - x_n)^2 + ((y_h - y_n)^2)}{z_n}\right)$. In order to reduce the usage of hardware resources, we set $a_n$ to 1. However, we can deal with arbitrary values of $a_n$ if we used an FPGA chip with larger resources. We can use the *popcount* technique (a.k.a Hamming weight) for Eq. (8), which can accelerate it to only $log_2(N)$ steps.

We have developed the binary CGH processor based on Eq. (8). Extended Data Figure 5(a) shows a conceptual scheme of this circuit. The processor computes all object points in parallel and consists of a "point-by-point calculation" (PPC) unit and a *popcount* unit. A PPC unit calculates a CGH pixel at $(x_h, y_h)$ from a single object point. In the case of $N$ object points, we prepare $N$ PPC units for each object point. Each PPC unit computes the same CGH coordinates, but computes different object points. In the current design, we can process the number of object points up to $N$ = 2048. Extended Data Figure 5(b) shows the timing chart of the processor. CGH pixels $I(x_h, y_h)$ can be computed in only one clock (the order of nanosecond) because each unit was fully-pipelined. Before invoking the processor, we set $N$ object point data $(x_j, y_j, z_j)$ to the memory in FPGA. The processor calculates other binary CGH pixels by feeding other CGH coordinates $(x_h, y_h)$ into the processor. We designed the PPC units with a 5-stage pipeline and the *popcount* unit with a 10-stage pipeline. Therefore, we need the latency of 15 clock cycles to obtain the first CGH pixel, but subsequent CGH pixels can be obtained within one clock cycle. It is worth noting that a single CGH pixel can be calculated by only one clock from $N$ object points, while conventional CPUs and GPUs take $N$ clock cycles; when $N$=2048, our processor achieved over 2000 times acceleration as compared to the traditional approach. In addition, the FPGA chip does not need external memory for holding CGH pixels, therefore the processor can be readily applied to CGH calculations with a high spatial bandwidth product. Equations (6) and (7) will be further reduced in hardware resources by using a recurrence algorithm, allowing one to treat more object points[43,44] in future iterations of the FPGA subsystem.

We have implemented the described binary CGH processor with 2048 PPC units and one *popcount* unit into a single FPGA chip of Virtex UltraScale+HBM FPGA (product part: xcvu35p-fsvh2104-3-e) made by Xilinx. The operation clock frequency was 125 MHz; thus, we could obtain one CGH pixel at 8 ns from 2048 object points. This can be scaled down to 1 ns by using modern FPGA chips with higher clock frequency up to 1 GHz without changing the proposed architecture, which was proven in FPGA simulations. The power consumption of the real FPGA CGH processor was about 14 W. The snapshot of a reconstructed movie from CGHs calculated by our processor is shown in Extended Data Fig. 6. The calculation conditions were that the CGH size of 1024×1024 pixels, the number of object points was 2048, the wavelength of 532 nm, and the distance of 0.3 m from the CGH. The input object was a 3D model of a moving dinosaur.

**Data and materials availability**

All experimental data shown in the main text and in the supplementary materials.

**Code availability**

The code is available from the corresponding authors upon reasonable request.

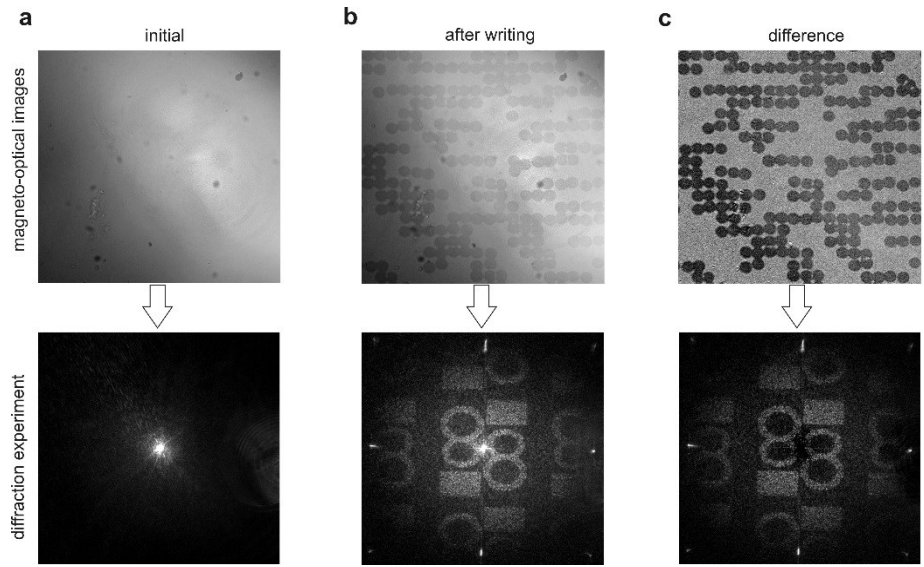

**Extended Data Fig. 1 | Opto-magnetic CGH writing and reading.** From left to right shows both the magneto-optical (top panel) and diffraction (bottom panel) images in the GdFeCo sample before the opto-magnetic writing (a), after writing (b), and the difference between these images (c). The magneto-optical images show the area of 200×200 µm².

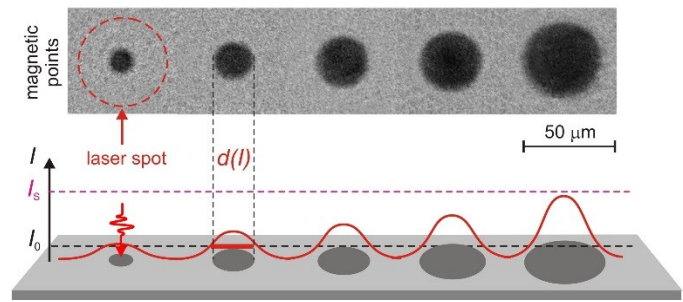

**Extended Data Fig. 2 | Writing single magnetic points using AOS.** The images of reversible writing by AOS magnetic points with a different pump fluence from left to right show: 12 mJ·cm$^{-2}$; 14.2 mJ·cm$^{-2}$; 15.6 mJ·cm$^{-2}$; 17 mJ·cm$^{-2}$ and 20 mJ·cm$^{-2}$. The horizontal dashed lines indicate the switching threshold and saturation of the laser pulse intensity *I*.

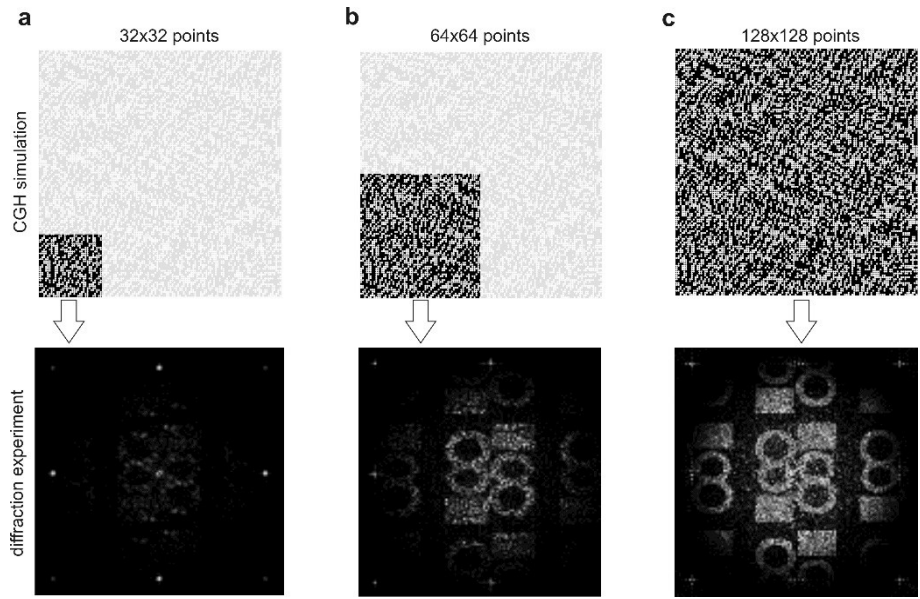

**Extended Data Fig. 3 |** Images of CGH pattern (top panel) and reconstructed in far field diffraction plane without division to frames in the time domain for points: (a) 32×32, (b) 64×64, (c) 128×182.

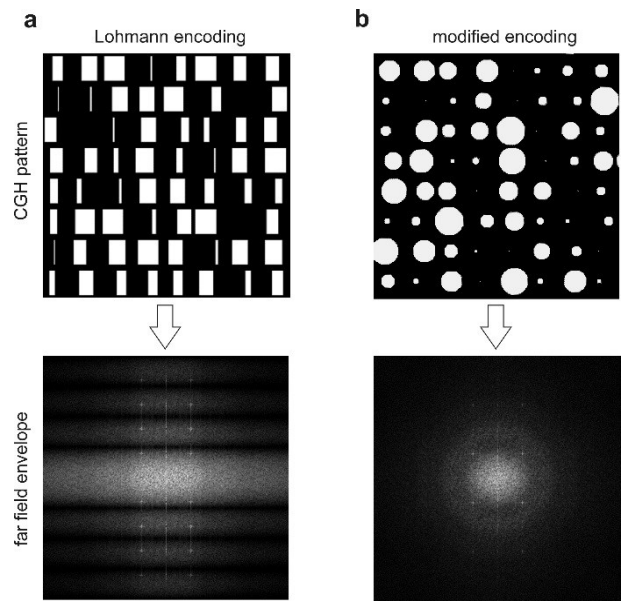

**Extended Data Fig. 4 | Complex encoding of CGH.** Shaping of the far field intensity envelope with modified Lohmann encoding comprising circular openings.

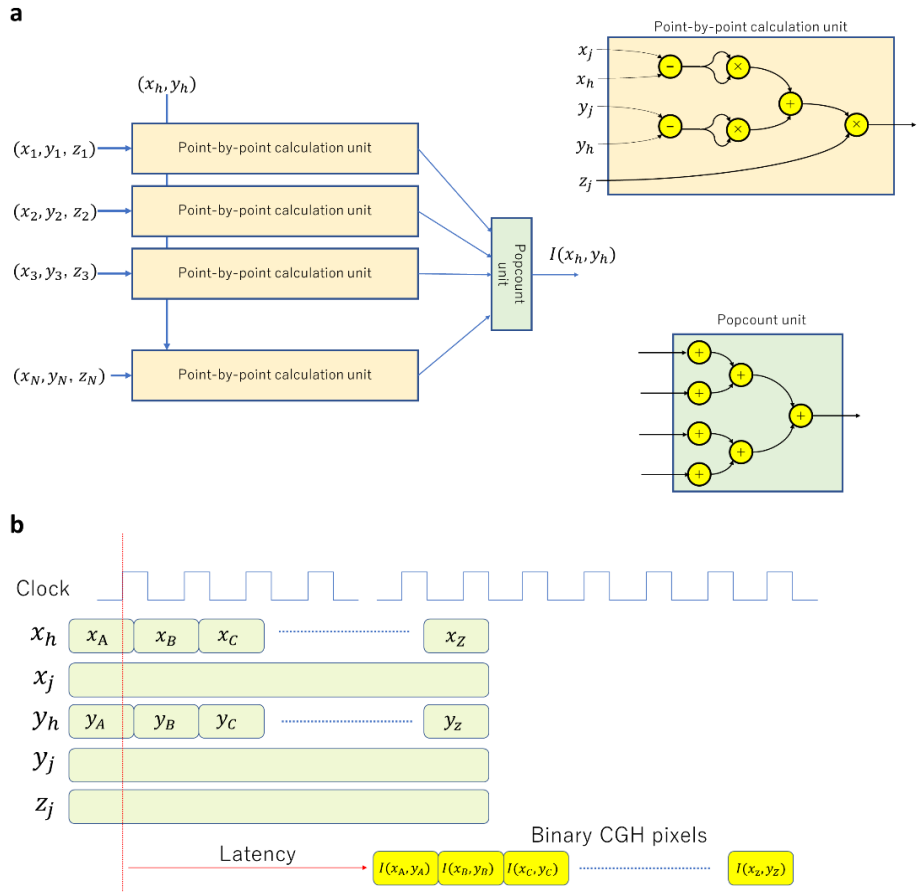

**Extended Data Fig. 5 | Serial computation CGH.** A binary CGH processor for opto-magnetic recording with *popcount* unit for the accumulation of four binary bits (a). Timing chart of the processor (b).

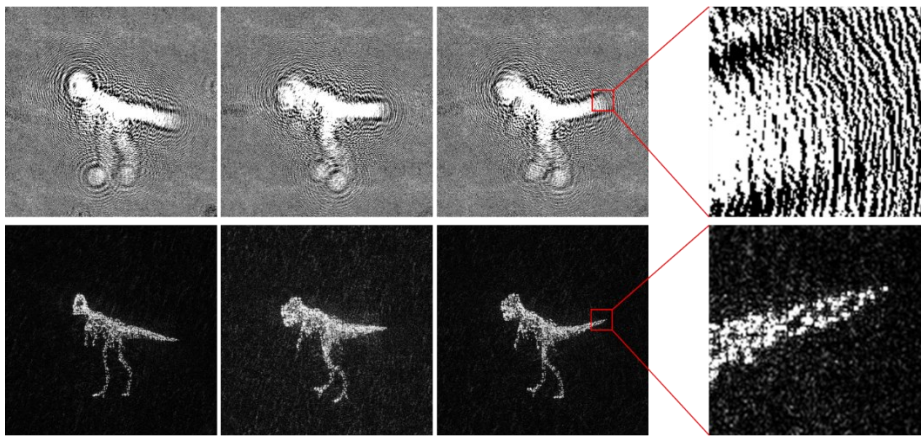

**Extended Data Fig. 6 |** Snapshot of a reconstructed movie from CGHs calculated by our processor. The top row shows the binary CGHs of the moving dinosaur, and the bottom row shows the corresponding reconstructed images.

**Extended Data Movie S1.** | The movie shows the dynamical sequence of opto-magnetic holographic reconstruction, demonstrating the counting between 0 and 9.